\documentclass[12pt]{article}
\usepackage{jccart}
\usepackage{graphicx}

\def\defpsfig#1#2#3#4#5#6{\begin{figure} \vspace{#2in}
    \hskip #6in\relax
    \ifusePS \includegraphics{#4} \fi
    \vspace{0.4in}
    \caption{\deffigcap{#1}{#3}}
    \end{figure}}

\orderrefs
\def \rf #1#2#3{#1 (19#3) #2}
\def \xbj {x_{{\rm Bj}}}
\def \qt {q_{\perp }}
\def \kt {k_{\perp }}
\def\TOp{{\cal T}}
\def\POp{{\cal P}}
\def\PT{\POp\TOp}

\def\secref#1{Sec.\ \OLDref{#1}}
\begin{document}
   \begin{flushright}
      \vskip 1.0in
        PSU/TH/102\\
        March 23,  1993 
      \\
   \end{flushright}
   \begin{center}
      \vskip 0.50in
      {\bf \large Fragmentation of Transversely Polarized Quarks
            Probed in Transverse Momentum Distributions}
      \vskip 0.1in
      {\bf John Collins \\}
      \vskip 0.3in
          Physics Department, Penn State University, \\
          {\it University Park, PA 16802, U.S.A.\\}
   \end{center}
   \vskip 0.3in
   \begin{quote}
      \begin {center}
               {\bf Abstract}
      \end {center}

      It is shown that the azimuthal dependence of the
      distribution of hadrons in a quark jet is a probe of the
      transverse spin of the quark initiating the jet.  This
      results in a new spin-dependent fragmentation function that
      acts at the twist-2 level.  One example of a process where
      it contributes is semi-inclusive deeply inelastic
      scattering with a transversely polarized nucleon target but
      with an {\em unpolarized} electron beam.  This process is
      treated in detail.  Another process is the cross section
      for singly polarized hadron-hadron scattering when two high
      $p_{\perp }$ hadrons are measured in the final state and are close
      to back-to-back in azimuth.  The new fragmentation function
      is sensitive to the coupling of the fragmentation process
      to (spontaneous) chiral symmetry breaking.

   \end{quote}

\section {Introduction}
\label{intro}

An important challenge to QCD theorists is to devise methods of
measuring the polarization state of a parton coming out of a hard
scattering.  In the case of longitudinal polarization for the
parton, Nachtmann \ref[Nacht] showed how a certain three-particle
correlation within a jet could be used.  He suggested several
processes where it could be measured, in particular deep
inelastic neutrino scattering. Later, Dalitz, Goldstein and
Marshall \ref[DMG] and Einhorn \ref[Einhorn] showed how to probe
the helicity of a heavy quark.

Recently Efremov, Mankiewicz and T\"ornqvist \ref[EMT]
rediscovered the Nachtmann idea, which they called the
`handedness' of a jet, and they showed how to measure it in
$e^{+}e^{-}$ annihilation. They also considered the possibility of
probing the {\it transverse} polarization of quarks.  This idea
was independently discovered in \ref[trfrag].

In this paper, I will present another technique sensitive to the
transverse polarization of a quark.  One measures the {\em
single} particle distribution as a function of transverse
momentum, in a process such as semi-inclusive deeply inelastic
scattering, where the jet axis can be precisely defined. The
initial-state hadron is transversely polarized, and the hard
scattering provides a spin transfer to the final state.  There
should be a spin asymmetry in the azimuthal distribution of the
outgoing hadron about the jet axis.

One could argue that the large number of hadrons produced in a
jet would completely dilute this asymmetry.  However, it is known
that for flavor quantum numbers, the leading particle in a jet is
correlated in flavor with the quark initiating the
jet \ref[EMCfrag].
Moreover, much of the hadron production results from the
emission of gluons at small $z$; this preserves the helicity state of the
quark, and it is the quark that is responsible for the leading
hadrons.

Measurements like this probe three quantities, none of which have
yet been measured: the transverse spin of a quark in a
transversely polarized hadron, the transverse spin dependence of
hard scattering, and the spin-dependence of fragmentation.  The
difference between distributions for the transverse spin and
for the helicity of a quark in
a hadron is a measure of relativistic effects in the hadron's
wave function \ref[JJ].  Transverse spin dependence of the
fragmentation involves a helicity flip for the quark, and is
therefore sensitive not only to the spin dependence of
fragmentation in general but to the presence of
(non-perturbative) chiral symmetry breaking.

For the single-particle measurement, one must have a process where
there is a definite axis or plane with respect to which
transverse momentum of an outgoing particle can be defined. Two
obvious examples are:
\begin{itemize}
\item Semi-inclusive deeply inelastic scattering,
$e+p^{\uparrow} \to e+H+X$, when the kinematic variables $\xbj$ and $z$
are not small, and the transverse momentum of the observed hadron
(relative to the plane of the initial proton and the two
electrons) is moderate.  This is the process that I will treat in
detail in this paper.
\item $p+p^{\uparrow} \to  H_{1}+H_{2}+X$, when the two observed
final-state hadrons have large $p_{T}$ relative to the axis of the
initial-state particles and are nearly back-to-back in azimuth
about that axis.
\end{itemize}
Here, $p^{\uparrow}$ denotes a transversely polarized hadron.
Undoubtedly, other similar observables can be easily devised.  It
is also standard to define a jet axis by calorimetric
measurements of the jet for example.  In that case one can
measure the azimuthal distribution of the leading pions about the
axis.  This probably suffers from lower precision in the
definition of the axis.

In the proton-proton case, if one makes
measurements at sufficiently large center-of-mass rapidity, then
one can have the collision of a small $x$ gluon and a moderately
large $x$  valence quark.  This allows one to take advantage of
the large number of small $x$ gluons and the expected large
polarization of the valence quarks. This would give both a large
spin asymmetry and a large cross section.

Both the single-particle measurement suggested in this paper and the
two-particle measurement suggested in \ref[EMT,trfrag] are twist-2:
that is, the spin asymmetries are not suppressed by a power of the
large momentum scale $Q$ in the hard scattering.  This is important
because of the paucity of useful twist-2 asymmetries involving
transverse spin.  The problem is that transverse spin asymmetries
involve off-diagonal elements in the helicity density matrices of the
quarks, and that the hard scattering preserves quark helicity.  (It is
quarks, not gluons that are relevant here \ref[AM].)  Thus to get a
twist-2 asymmetry, the non-perturbative part of the process
(consisting of the parton distribution and fragmentation functions)
must contain an even number of factors that flip quark helicity.

The obvious measurements that are sensitive to transverse spin
at the twist-2 level involve collisions of {\em two}
polarized hadrons, and to get a large asymmetry one wants to be
in the valence region for the quarks entering the hard
scattering.  The two standard processes are jet production and
the Drell-Yan process.  But jet production has a numerically
small asymmetry \ref[trjet] in the hard scattering. The Drell-Yan
process, with a large asymmetry in the hard scattering \ref[RS],
needs a polarized antiproton beam with good luminosity to make it
a valence process with a reasonable event rate.

In contrast, the spin asymmetries discussed in this paper, and
those in  \ref[EMT,trfrag] have their second measured spin in the
final state instead of the initial state.  It is interesting that
these fragmentation measurements are easier for transverse
polarization than for longitudinal.  A transverse polarization
can be probed by measuring two suitable momentum vectors (e.g.,
two particles in a jet, or one particle and a jet axis).  But a
longitudinal polarization requires three vectors, to define a
handedness---see \ref[Nacht,EMT], and also the work of Carlitz
and Willey \ref[CarlitzDY].

The new spin dependent fragmentation effect that is the
subject of this paper gives a leading-twist contribution
only to processes that are sensitive to intrinsic transverse
momentum.  But it may also have consequences for other processes.
For example, consider the inclusive cross section for production
of a high $p_{\perp }$ hadron in hadron-hadron collisions.  When only one
of the initial hadrons is transversely polarized, QCD predicts
that the resulting spin asymmetry is of twist-3.
There is a notorious contrast between
this prediction and the large asymmetry actually
measured \ref[pion].  The experiments are admittedly at rather
modest energies (fixed target with beams of up to 200 GeV).
Methods for treating twist-3 effects are currently begin
developed \ref[JJ,BB,QS].

In that process, there is effectively an average over intrinsic
transverse momentum, and in particular over its azimuthal angle
about the jet axis. Therefore the spin-dependence of the
fragmentation that is treated in the present paper gets
canceled, at the leading twist level.  But the cross section is a
steeply falling function of the hadron's $p_{\perp }$, so that the
azimuthal average need not work very well.  The result could be a
numerically large twist-3 contribution in the fragmentation.  The
authors of \ref[QS] have not yet treated twist 3 effects in
fragmentation.

The layout of this paper is as follows.  I define the
fragmentation and distribution functions in \secref{distfrag}.
Then in \secref{dis} I show how these get used in deeply
inelastic scattering, and in \secref{dispred} I explain the
polarization sensitive measurements that can be made; that
section is perhaps the one of most use to experimentalists.  In
\secref{hadhad}, I make a few remarks about hadron-hadron
scattering.  Then I exhibit
some simple model calculations in \secref{model} to show that the
fragmentation asymmetry I propose does not violate any
fundamental principles.
I summarize my conclusions in \secref{concl}.

\section {Spin-Dependence of Parton Distribution and Fragmentation
Functions}
\label{distfrag}

In this section, I will first review the formal definitions of the
parton distribution and fragmentation functions when there is a
measured transverse momentum.  Then I will show how to extend the
definitions to treat nontrivial polarization.  These quantities
will get used in factorization formulae for the cross section, as
explained in later sections.

In the usual factorization theorems \ref[CSS,CSS1], one works with
parton densities integrated over transverse momentum.  But when
one has a cross section with a measured small transverse momentum
variable, one must use the unintegrated distributions.  In QCD,
there are some interesting effects associated with Sudakov form
factors, that make the resulting factorization theorems quite
nontrivial \ref[CSS2].  The Sudakov effects are spin-independent,
and we will not bother making them explicit here, since our
purpose is to examine the novel effects associated with
polarization.  However when the energy of the experiment
increases, the Sudakov effects will dilute our asymmetries
by smearing out the transverse momentum distributions.

We will denote the unpolarized distribution and
fragmentation functions by $f_{i/H}$ and $D_{H/i}$
respectively, when transverse momentum is integrated over.  To
denote the corresponding quantities with unintegrated transverse
momentum, we will use the same symbols, but with a hat over them:
$\hat f_{i/H}$ and $\hat D_{H/i}$.

\subsection {Parton Distribution Functions}

We define parton distribution functions by formulae motivated by
light-front quantization.  These quantities are precisely those
that occur in the factorization theorems \ref[CS,DYJCC,LC].

It follows from the parity and time-reversal invariance of QCD
that the number density of quarks is independent of the spin
state of the initial hadron, so that we have
$$
   \hat f_{a/A}(x,|\kt|) \equiv
   \int  \frac {\d y^{-} \d^{2}y_{\perp }}{(2\pi )^{3}}
   \e^{-ixp^{+}y^{-}+i\kt\cdot y_{\perp }}
   \langle p| \, \bar \psi _{i}(0,y^{-},y_{\perp }) \,
   \frac {\gamma ^{+}}{2} \, \psi _{i}(0) \, |p\rangle .
\eqno(1)
$$
We have ignored here the subtleties needed to make this a gauge
invariant definition: an appropriate path ordered exponential of
the gluon field is needed \ref[CS].  The coordinate frame in which
this definition is applied is one in which the hadron $|p\rangle $ has
zero transverse momentum: $p_{\perp }=0$.

Sivers \ref[Sivers] suggested that the $\kt$ distribution of the
quark could have an azimuthal asymmetry when the initial hadron
has transverse polarization.  However, such an asymmetry is
prohibited because QCD is time-reversal invariant.  This is
shown in the appendix.

As explained in \ref[RS,polfact], we must consider the quark (or
gluon) $a$ to be equipped with a helicity density matrix.  Since
QCD is invariant under parity and time reversal, the density
matrix for a quark
differs from unity only if the initial hadron $A$ is
itself polarized.   Then the transverse spin asymmetry of a quark is
defined by:
$$
\eqalign{
   s_{\perp }^{\mu }\hat \Delta _{Ta/A}\hat f_{a/A}(x,\kt)
   &\equiv  s_{\perp }^{\mu }\hat f_{Ta/A}(x,\kt)
\cr
   &\equiv
   \int \frac {\d y^{-} \d^{2}y_{\perp }}{(2\pi )^{3}}
      \e^{-ixp^{+}y^{-}+i\kt\cdot y_{\perp }}
   \langle p| \, \bar \psi _{a}(0,y^{-},y_{\perp }) \,
      \frac {\gamma ^{+}\gamma _{\perp }^{\mu }\gamma _{5}}{2} \,
   \psi _{a}(0) \, |p\rangle ,
}
\eqno(2)
$$
where $s_{\perp }^{\mu }$ is the transverse part of the spin vector of
the initial hadron, normalized so that its maximum
size is unity: $|s_{\perp }| \le 1$.  In \eq(2), I have used
the notation of \ref[polfact], where definitions are given for the
case that the transverse momentum is integrated over; the
definitions given for that case given by Jaffe and Ji \ref[JJ]
differ only in notation.  I have defined $\hat\Delta _{T}$ to mean the
ratio of quark polarization to hadron polarization; it is a
kind of asymmetry or spin transfer function, and in general will
depend on $x$ and $\kt$.  Then $\hat f_{T}$ (with a subscript $T$)
means the parton distribution weighted by the transverse
spin asymmetry.

Similar definitions can be given for the distribution of gluons.
It would be interesting to work out the details.  For the
deep-inelastic process treated in this paper, we will see that we
will not need the definitions for gluons.

One can also write helicity asymmetries. But we will not need
them in this paper, because we will work with fragmentation
observables that are not sensitive to quark helicity.

\subsection {Fragmentation Functions}

Fragmentation functions with transverse momentum are defined
in a similar fashion to the parton distribution functions.
An important difference is that the observed final-state hadron,
assumed to be a pion, has no polarization, so that a
spin-transfer function analogous to \eq(2) does not exist.  On
the other hand we can no longer apply the generalization of the
time-reversal argument that prevented an azimuthal asymmetry in
the number density, as we will explain.

The unpolarized fragmentation function, to find a hadron $H$ in
the decay products of a quark of flavor $a$ is \ref[CS]
$$
  \hat D_{H/a}(z,\kt) \equiv
    \sum _{X} \int \frac {dy^{-}d^{2}y_{\perp }}{12 (2\pi )^{3}}
        \e^{ik^{+}y^{-}-i\kt \cdot y_{\perp }}
     \tr \gamma ^{+} \langle 0| \psi _{a}(0,y^{-},y_{\perp }) |HX\rangle
      \langle HX| \bar \psi _{a}(0) |0\rangle .
\eqno(3)
$$
The factor of $1/12$ is the product of a factor $1/2$ which
occurs with all these definitions for fermions and a factor $1/6$
for an average over the spin and colors states of the quark. The
sum over $X$ is over all final states containing the chosen hadron.

This definition is easily generalized to give the transverse spin
dependence $\Delta \hat D$ of the distribution of hadrons in a
polarized quark of transverse spin $s_{\perp }^{\mu }$:
$$\eqalign{
   \Delta \hat D_{H/a}(z,\kt ,s_{\perp }) \equiv  &
\cr
  \sum _{X} \int  & \frac {dy^{-}d^{2}y_{\perp }}{12 (2\pi )^{3}}
    \e^{ik^{+}y^{-}-i\kt \cdot y_{\perp }}
    \tr \gamma ^{+}\gamma _{5}\, \gamma _{\perp }{\cdot }s_{\perp }
    \langle 0| \psi _{a}(0,y^{-},y_{\perp }) |HX\rangle
     \langle HX| \bar \psi _{a}(0) |0\rangle .
}\eqno(4)
$$
This is permitted to depend on the spin through the following
scalar quantity:
$$
  \epsilon _{\kappa \lambda \mu \nu }
  s_{\perp }^{\kappa } \kt ^{\lambda } p_{H}^{\mu } n^{\nu },
\eqno(5)
$$
where $n^{\mu }\equiv \delta ^{\mu }_{-}$ is the vector that is
used to define the
light-front momentum fraction $z$:
$z\equiv p_{H}^{+}/k^{+}=p_{H}\cdot n/k\cdot n$, with $k^{\mu }$
being the momentum of the quark. (Note that the above definitions
imply that $k^{-}$ is integrated over, while $\kt$ is the transverse
momentum of the quark relative to the hadron $H$.)

There is some disagreement in the literature on the normalization
of the fragmentation functions.  The definitions given above are
the most convenient for theoretical analyses since the bilocal
vertex for the quark depends only on momenta defined at that
vertex.  However, a probability interpretation is most naturally
given in terms of the transverse momentum of the hadron relative
to the quark, which is $p_{\perp }=-z\kt$.  Then the probability density
in $z$ and $p_{\perp }$ is $z^{-1}\hat D(z,-z p_{\perp })$.
As for the density
integrated over transverse momentum,
i.e., $D(z) \equiv  \int d^{2}\kt \hat D(z,\kt)$,
the probability density is $zD(z)$.  (This implies
that the momentum sum rule is
$\sum _{H}\int _{0}^{1}z^{2}D_{H/a}(z)\,dz = 1$.)

Now time reversal invariance prohibits an asymmetry proportional
to \er(5), unless there are nontrivial phases from final state
interactions; but such phases can surely exist in a strong
interaction fragmentation.  Another way of saying this comes from
observing that a combination of time reversal and parity
invariance transforms out-states to in-states of the same
momentum content.  (Compare the arguments given in the
appendix for the distribution functions.)
If the in-states were the same as the
out-states, then we could show that the asymmetry
proportional to \er(5) is zero.
For strongly nonperturbative interactions, such as are involved
in hadronization, this is surely not true.
Similar arguments
have been applied to polarized pion-nucleon scattering \ref[Gas].

\section {Semi-Inclusive Deeply Inelastic Scattering}
\label{dis}

\subsection {Kinematics}

Semi-inclusive deeply inelastic scattering is the process
$e+A \to  e'+B+X$ where $A$ is some arbitrary initial hadronic state
of momentum $p_{A}^{\mu }$, and $B$ is an observed hadron in the final
state of momentum $p_{B}^{\mu }$.  The momentum transfer from the lepton
system is $q^{\mu }$, as usual, and we define the masses of hadrons $A$
and $B$ to be $M_{A}$ and $M_{B}$.

In the absence of polarization, there are four Lorentz invariant
variables for the hadronic system, viz.,
$q^{2}$, $p_{A}\cdot q$, $p_{B}\cdot q$, and $p_{A}\cdot p_{B}$.
A useful set of
variables was defined by Meng, Olness and Soper \ref[MOS].  The
two variables not involving the final-state hadron $B$ are, as
usual,
$$
   \xbj \equiv  \frac {Q^{2}}{2p_{A}\cdot q},    \eqno(6)
$$
and
$$
   Q \equiv  \sqrt {-q^{2}}.     \eqno(7)
$$
Then as in \ref[F] one defines
$$
   z \equiv  \frac {p_{A}\cdot p_{B}}{p_{A}\cdot q}.      \eqno(8)
$$
Finally, Meng et al.\ define $\qt^{\mu }$ as the transverse momentum of
$q^{\mu }$, in a frame where $p_{A}^{\mu }$ and $p_{B}^{\mu }$ have
zero transverse
momentum. Up to corrections that are irrelevant in the Bjorken
limit, this means that
$$
\qt^{\mu } = q^{\mu } + \bar x_{A} p_{A}^{\mu } - \frac {1}{z}p_{B}^{\mu },
\eqno(9)
$$
where $\bar x_{A} \equiv  -p_{B}\cdot q/p_{A}\cdot p_{B} =
\xbj \left(1-|\qt^{2}|/Q^{2} \right)$.

The deep inelastic region is where $Q$ is made large, with $\xbj$
and $z$ held fixed and not close to their endpoints 0 and 1.  We
will always assume in this paper that the scattering is taken to
lowest order in QED, with a single photon being exchanged between
the lepton and the hadronic system, \fig{1}.

   \defpsfig{1}{2}{Kinematics of semi-inclusive deeply inelastic
   scattering.  }{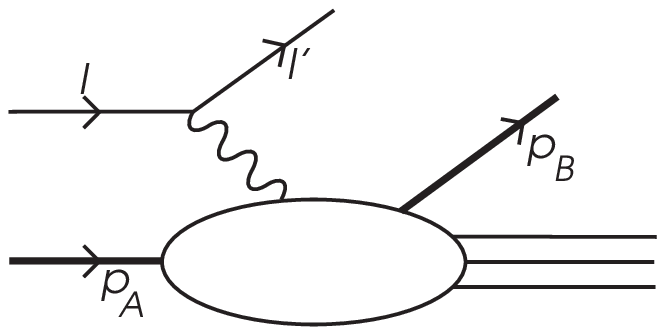}{1000}{2.0}

The reason for defining the variables $\xbj$, $z$ and $\qt$ is
that they have a simple interpretation in the parton model.
There, it is assumed that the dominant contributions to the cross
section have the form of \fig{2}.  The virtual photon interacts
in Born approximation with a single quark, which is close to its
mass shell and which has low transverse momentum on the scale $Q$.
Then when hadron $B$ is part of the `current quark
jet' produced in the hard scattering, $z$ has the interpretation
of the fraction of the jet's momentum that is carried by the
hadron.  As usual, $\xbj$ has the interpretation of the fraction
of the momentum of the incoming hadron $A$ that is carried by the
parton that enters the hard scattering.

   \defpsfig{2}{3}{Parton model for semi-inclusive deeply inelastic
   scattering.  }{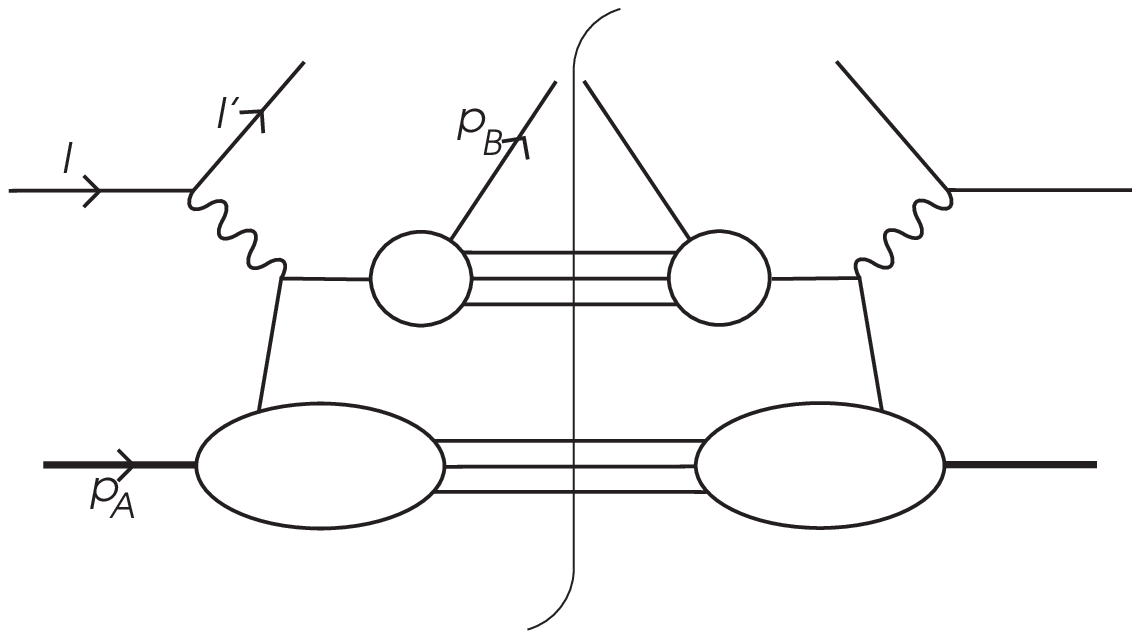}{1000}{1.0}

To treat intrinsic transverse
momentum for the initial state and for the fragmentation, we need
a suitable frame in which to define them.  First we
define what we will call the `parton model jet axis':
$$
  p_{J}^{\mu } \equiv  q^{\mu } + \xbj \, p_{A}^{\mu }.
\eqno(10)
$$
This would be the jet momentum if there were no intrinsic
transverse momentum.  Even in the presence of transverse
momentum, the definition \er(10) gives a convenient axis with
respect to which to define the transverse momentum of the
outgoing hadron $B$.

As stated above, $\qt$ is the transverse momentum of the virtual
photon in a frame in which the two hadrons $A$ and $B$ have zero
transverse momentum.  This is perhaps the simplest frame in which
to derive the factorization theorem at low transverse momentum.
But when we consider the azimuthal distribution of the outgoing
hadron $B$ it will be more intuitive to consider the transverse
momentum of $B$ relative to the parton model jet axis.  This we
can define by
$$
   p_{BT}^{\mu } \equiv  p_{B}^{\mu } - \frac {1}{z}p_{J}^{\mu }.
\eqno(11)
$$
Note that to avoid some possible confusion we use the subscript
$_{T}$ here rather than $_{\perp }$.
This is to emphasize that $p_{BT}^{\mu }$ is a
four vector rather than the transverse two-component vector
defined in a particular frame: the frame in which we define the
fragmentation function will have zero transverse momentum for
$B$.

Up to corrections of relative order $\qt^{2}/Q^{2}$, the transverse
momentum $p_{BT}^{\mu }$ is $-z\qt^{\mu }$.

\subsection {Factorization without intrinsic transverse momentum}

In QCD, there is a standard factorization theorem:
$$
   E'E_{B}\frac {d\sigma }{d^{3}l' \, d^{3}p_{B}}
   = \sum _{a,b}\int d\xi  \int \frac {d\zeta }{\zeta } \,
     f_{a/A}(\xi ) \, (\hbox{Hard scattering}) \,D_{B/b}(\zeta ),
\eqno(12)
$$
which is correct up to power-law corrections. When the hard
scattering factor is replaced by its lowest order approximation,
we regain the parton model.  In normalization the hard scattering
function is the invariant cross section for lepton-parton
scattering with the outgoing parton momentum set to $p_{B}/\zeta $.

This factorization theorem represents the asymptotics of the
cross section when $Q$ gets large but the dimensionless ratios $\xbj$,
$z$ are held fixed, and $\qt/Q$ is either held fixed or integrated over.
This has the unfortunate implication that the factorization
theorem fails to give a useful result for the distribution in
$\qt$ when $\qt \ll Q$.  However, this is exactly where the cross
section is biggest.  Indeed, in lowest order, the hard scattering
function has a factor $\delta ^{(2)}(\qt)$, while in higher order
the hard scattering function has terms that diverge like $1/\qt^{2}$
times logarithms when $\qt \to 0$.

A convenient way of formulating these statements is to state that the
factorization theorem \eq(12) must be interpreted in the sense of
distribution theory \ref[Tk]. The theorem applies when the cross
section is integrated with a smooth test function
$\phi (\xbj,z,\qt/Q)$.  In the derivation \ref[CSS,CSS1,polfact], one
applies to the hard scattering the approximation that the transverse
momentum of the incoming parton can be neglected with respect to the
transverse momentum generated in the scattering, and one also
neglects the transverse momentum generated in the fragmentation.
(Note that to the extent that these transverse momenta are not
negligible, but are of order $Q$, the errors in the
approximations are compensated by a correct treatment of higher
order corrections to the hard scattering.)

\subsection {Factorization with intrinsic transverse momentum but no
  polarization}

To gain information on the $\qt$ dependence at small $\qt$,
we must derive a more
powerful theorem that involves `intrinsic transverse momentum' in
both the distribution and the fragmentation functions.
Such a theorem was derived for the Drell-Yan
process and for the two-particle-inclusive cross section in
$e^{+}e^{-}$ annihilation \ref[CSback,CSS2].  A similar theorem should apply
here.  An obvious ansatz is
$$
\eqalign{
  E'E_{B}\frac {d\sigma }{d^{3}l' \, d^{3}p_{B}} = &
  \sum _{a} \int d\xi  \int \frac {d\zeta }{\zeta } \int d^{2}k_{a\perp }
  \int  d^{2}k_{b\perp }
  \, \hat f_{a/A}(\xi ,k_{a\perp })
  \,\,
  \, E'E_{k_{b}}\frac {d\hat\sigma }{d^{3}l' \, d^{3}k_{b}}
  \, \hat D_{B/a}(\zeta ,k_{b\perp })
\cr
  & + Y(\xbj,Q,z,\qt/Q).
}
\eqno(13)
$$
In this formula $\hat\sigma $ represents the short distance part of
elastic lepton-quark scattering.  It contains
a delta-function for momentum conservation.  The sum over $a$ is
over all flavors of quark and antiquark.

The first term on the right of \eq(13)
dominates when $\qt \ll Q$.  The second term, $Y$, is a correction
term that enables \eq(13) to reproduce the ordinary
factorization theorem \eq(12) at large transverse momentum, just
as in the Drell-Yan case \ref[CSS2].  The $Y$ term has the general
form of the basic factorization theorem \eq(12), except that the
low $\qt$ asymptote is subtracted from the hard scattering
function.

The function $\hat f_{a/A}$ defined earlier gives
the intrinsic transverse momentum dependence of partons in the
initial state hadron.
Similarly, $\hat D_{B/a}$ gives the distribution of hadrons in a
parton, with $k_{b\perp }$ being the transverse momentum of the parton
relative to the hadron.

Just as in \ref[CSback,CSS2], the hard scattering factor in the first
term in \eq(13) can only be a $2\to 2$ process. Hence the fractional
momenta of the incoming quark from hadron $A$ and of hadron $B$
in the outgoing quark are forced to be $\xbj$ and $z$.

After integrating out the delta-function in $\hat\sigma $ we obtain
$$
\eqalign{
  E'E_{B}\frac {d\sigma }{d^{3}l' \, d^{3}p_{B}} = &
  \frac {4\xbj}{Q^{2}} \sum _{a} \int d^{2}k_{a\perp }
  \, \hat f_{a/A}(\xbj,k_{a\perp })
  \,\,
  \, \frac {d\hat\sigma }{d\Omega }
  \, \hat D_{B/a}(z,k_{a\perp }+\qt)
\cr
  & + Y(\xbj,Q,z,\qt/Q).
}
\eqno(14)
$$

The picture that goes with these results is \fig{2}.  All that we
have done is to take account of the transverse momentum of the
quarks relative to the measured initial- and final-state hadrons.
This intrinsic transverse momentum has the effect of smearing out
the delta function of $\qt$ that we remarked on earlier.  The
only generalization needed compared with the parton model is that
the hard scattering can contain higher order {\em virtual}
corrections.  In the absence of gauge bosons in the strong
interactions, this formula in the exact form given in \eq(13) is
a theorem, that can be proved as in the Drell-Yan
case \ref[DYJCC].

The spin-1 gluons of QCD modify the theorem, by causing Sudakov
form-factor effects.  We expect that a proof can be given just as
in the Drell-Yan case \ref[CSS2].  The effect is to broaden the
transverse momentum distribution as $Q$ increases, but in a
spin-independent way: the broadening is due to recoil against the
transverse momentum of soft gluon emission.  This will have the
effect of diluting the spin asymmetry we will discuss next.

\subsection {Factorization with intrinsic transverse momentum and
   polarization}

We now explain factorization for the semi-inclusive deep
inelastic cross section when the incoming hadron $A$ is
transversely polarized but the lepton remains unpolarized.
(It is left as an exercise to treat the most general case.) The
factorization theorems, \eq(12) and \eq(14), continue to apply
when we include polarization for the incoming hadron, but with
the insertion of
helicity density matrices for in and out quarks; this is a
simple generalization of the results in \ref[RS,polfact].

The cross section will be linear in the transversity
$s_{\perp }^{\mu }$ of the
hadron (and also linear in its helicity $\lambda $). Because transverse
spin for a spin-\half{} particle corresponds to off diagonal
terms in the helicity density matrix, the other primary
constraint comes from quark helicity conservation in the hard
scattering, and this simplifies the factorization theorem.

First, it is well known that at large transverse momentum, the
transverse spin asymmetry is higher twist, as I now review.
In that region, we use distribution and fragmentation functions
integrated over intrinsic transverse momentum.  Now, in the
absence of a measurement of the polarization of the outgoing
hadron, the single-particle fragmentation is spin independent.
On the other hand, the transverse-spin dependence of the
distribution functions is only in the off-diagonal elements of
quark density matrices \ref[AM]. Therefore we need the part of the
hard scattering that is off-diagonal in the helicity of the
initial state quark but diagonal in the (summed) final-state
helicities. Helicity conservation at the vertices for the gluon,
photon and $Z$ prohibits such a term, at leading twist.

But, at low transverse momentum, the fragmentation function has
dependence on transverse spin --- see \eq(4).  The corresponding
hard scattering is just elastic electron-quark scattering, and we
need terms that are off-diagonal in the final state quark
helicity.  Quark helicity conservation requires that these terms
are also off-diagonal in initial-state quark helicity, and we are
therefore discussing a spin transfer from the incoming to the
outgoing quark.  The off-diagonal terms in the density matrix for
the incoming quark are given by the transverse spin distribution
defined in \eq(2).

Since there is a kinematic zero in the transverse spin dependence
of the fragmentation at $\kt=0$, the spin dependence of the cross
section must have transverse momentum dependence roughly
proportional to the dimensionless coefficient
$$
   \frac {\qt\, M}{\qt^{2}+M^{2}},
\eqno(15)
$$
where $M$ is a typical hadronic mass.  This exhibits the
kinematic zero when $\qt=0$, the leading twist asymmetry when
$\qt = O(M)$, and the higher twist asymmetry when $\qt \gg M$.

We have now seen that the only transverse spin asymmetry at
leading twist is in the low transverse momentum term in the
generalized \eq(14).  Moreover \er(5) implies that the transverse
spin asymmetry has a $\sin\phi $
dependence on the azimuth $\phi $ of the
transverse momentum, so that any kind of uniform azimuthal
averaging will remove the asymmetry.

Following Meng, Olness and Soper \ref[MOS] and earlier work, we
could decompose the cross section in terms of scalar structure
functions.  In the case of fully inclusive unpolarized deeply
inelastic scattering, with photon exchange, there are just two
structure functions, the well-known $F_{1}$ and $F_{2}$.  But when we
measure one of the particles, $B$, in the final state, there is
an extra vector in the problem, so there are more structure
functions, enumerated in \ref[MOS].  When in addition we allow
the incoming hadron to be polarized, there are extra structure
functions, just as for the kinematically isomorphic Drell-Yan
cross section \ref[RS,DG].
We will choose instead to work directly with the cross section
and its angular dependence.  It would of course be of interest to
perform a structure function analysis.

Since at leading twist, there is only transverse spin dependence in the low
transverse momentum term in \eq(14), it is only this term that we
need to change, with the result
$$
\eqalign{
 E'E_{B}\frac {d\sigma }{d^{3}l' \, d^{3}p_{B}} = &
 \frac {4\xbj}{Q^{2}} \sum _{a} \int d^{2}k_{a\perp }
 \, \hat f_{a/A}(\xbj,k_{a\perp })
 \, \rho _{\alpha \alpha '}
 \,\,
 \,\frac {d\hat\sigma _{\alpha \alpha ';\beta \beta '}}{d\Omega }
 \, \hat D_{\beta \beta ';B/a}(z,k_{a\perp }+\qt)
\cr
 & + Y(\xbj,Q,z,\qt/Q).
}
\eqno(16)
$$
Temporarily we have changed notation:
$\alpha $, $\alpha '$, $\beta $ and $\beta '$
represent indices for the initial and final-state densities
matrices, $\rho _{\alpha \alpha '}$ is the helicity
density matrix for the incoming
quark, while the indices $\beta \beta '$ on the fragmentation function
$\hat D$ represent the dependence of the fragmentation on the
outgoing quark's density matrix.

We now recast this formula in terms of transverse spin vectors by
using parity invariance as well as helicity conservation.  First,
we have already seen that helicity conservation tells us that the
only transverse spin dependence is in the part of the hard
scattering that is off-diagonal in helicity, i.e.,
$\hat\sigma _{+-;+-}$
and its hermitian conjugate $\hat\sigma _{-+;-+}$.  Thus the off-diagonal
terms in the final and initial state density matrices are
proportional:
$\rho ^{{\rm out}}_{+-} = C \rho _{+-}$, where $C$ depends on the
kinematic variables of the hard scattering.  Furthermore, the
expression for the off-diagonal term in a density matrix in terms
of the spin vector is
$\rho _{+-}= \half |s_{\perp }|\e^{i\chi }$, apart from a
possible phase and a possible sign error in the exponent.  Here
$\chi $ is the azimuthal angle of the transversity about the momentum
of the quark. Furthermore parity invariance implies that if the
spin of the initial quark is perpendicular to the plane of the
scattering, then so is the spin of the final-state quark.  Hence
the coefficient $C$ is real, if we choose our conventions such
that a real value of $\rho _{+-}$ corresponds to a spin vector
perpendicular to the plane of scattering.

It follows that the transversities of the initial and final
quarks are proportional.  The direction of the transversity of
the final quark is obtained by rotating it about the normal to
the plane of the hard scattering.  (\fig{3}.)

   \defpsfig{3}{2.6}{Relation between spin vectors for initial and
   final state quarks.  }{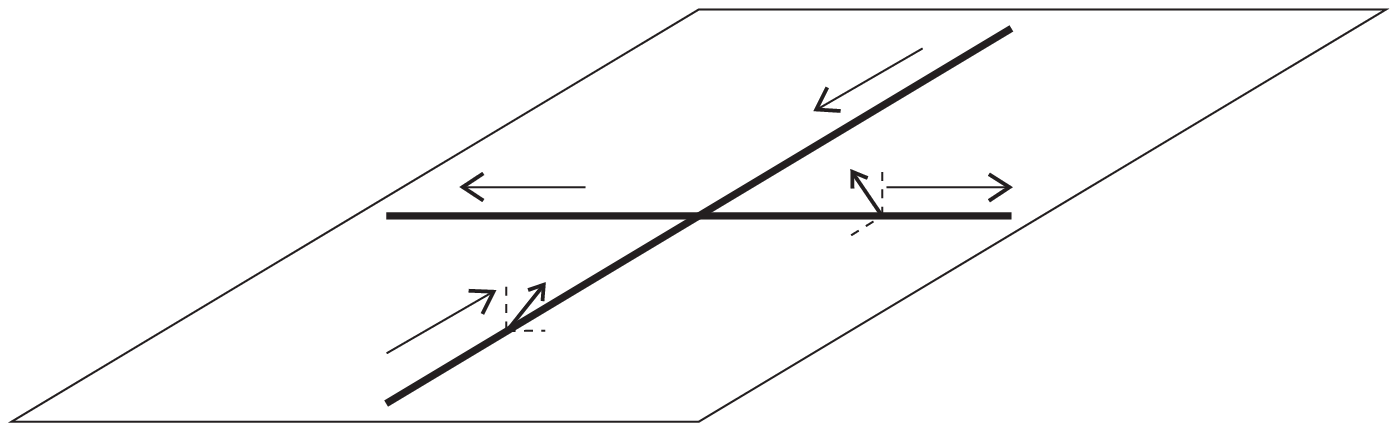}{1000}{0.7}

Hence the spin dependence of low transverse momentum term in the
cross section is given by the following factor:
$$
   1 + C
   \frac {\sum _{a} e_{a}^{2} \int d^{2}k_{a\perp } \Delta _{T}\hat f \Delta
\hat D(s')}{\sum _{a} e_{a}^{2} \int d^{2}k_{a\perp } \hat f \hat D} .
\eqno(17)
$$
Here we have used the fact that the spin transfer coefficient
$C$, which we will calculate in the next section, depends only on
the scattering angle of the hard scattering.
It is independent of quark flavor and the
other kinematic variables; thus we can factor it out of the
integrals. The sums are over flavors of quark and antiquark. The
transversity vector $s'$ that is used in the spin-dependent
fragmentation function $\Delta \hat D(s')$ is the transverse spin
vector of the initial hadron rotated in the center of mass of the
hard scattering, according to the method of \fig{3}.

The initial state distributions are functions of the length of
$k_{a\perp }$ only.  So after integration over
$k_{a\perp }$, the cross section
acquires a spin dependence whose dependence on the azimuth of $B$
is proportional to $\sin \phi $, where $\phi $ is the azimuthal angle of
$p_{B}$, when measured in the lepton-parton center-of-mass
about the outgoing jet axis (as defined by $p_{J}^{\mu }$ in
\eq(10)), with $\phi =0$
being the direction of the vector $s'$ defined just above.

\subsection{Calculation of spin transfer}

We can easily compute the spin transfer part
of the hard scattering cross section from the
amplitudes given by Gastmans and Wu \ref[GW].  The cross section
is for elastic electron-quark scattering, with all masses set to
zero.  Because of helicity conservation and parity invariance,
there are two independent helicity amplitudes: $M_{++}$ and $M_{+-}$,
where the first and second indices label the helicities of the
electron and quark.  The amplitudes $M_{--}$ and $M_{-+}$ are related
to these by parity invariance.

The unpolarized cross section is a standard kinematic factor times
$\frac {1}{2} \left[ |M_{++}|^{2} + |M_{+-}|^{2} \right]$.
When the electron is unpolarized, the
density matrix $\rho ^{f}$ of the outgoing quark is expressed in terms
of the density matrix $\rho ^{i}$ of the incoming quark as follows:
$$
\rho ^{f} \equiv
  \left( \begin{array}{ll}
           \rho ^{f}_{++} & \rho ^{f}_{+-} \\
           \rho ^{f}_{-+} & \rho ^{f}_{--}
         \end{array}
  \right)
=
  \left( \begin{array}{ll}
           \rho ^{i}_{++} & C\rho ^{i}_{+-} \\
           C\rho ^{i}_{-+} & \rho ^{i}_{--}
         \end{array}
  \right) ,
\eqno(18)
$$
where
$$
  C =
  \frac {M_{++}M^{*}_{+-} + M_{-+}M^{*}_{--}}{|M_{++}|^{2} + |M_{+-}|^{2}} .
\eqno(19)
$$
Up to a phase convention, the explicit values for the matrix
elements are \ref[GW]:
$$
\eqalign {
   M_{++} = M_{--}
   &= 4 i e^{2}q_{a} \frac {1}{1-\cos \theta } ,
\cr
   M_{-+} = M_{+-}
   &= 2 i e^{2}q_{a} \frac {1+\cos \theta }{1-\cos \theta } .
}
\eqno(20)
$$
Here $q_{a}$ is the electric charge of quark $a$ in units of the
basic charge $e$, while $\theta $ is the scattering angle in the
lepton-quark center-of-mass.  The definitions of the helicity
states are such that the scattering amplitude is purely
imaginary; this convention differs from that of \ref[GW].

It follows that the spin transfer coefficient is
$$
\eqalign{
  C &= \frac {4 (1 + \cos \theta )}{4 + (1 + \cos \theta )^{2}}
\cr
    &= \frac {1-y}{1-y+\frac {1}{2}y^{2}} ,
}
\eqno(21)
$$
where $y$ is the usual variable
$p_{A}\cdot (l-l')/p_{A}\cdot l$, which
is $(E-E')/E$ in the target's rest frame and $(1-\cos\theta )/2$ in the
lepton-quark center-of-mass.  The spin transfer coefficient is
100\% at small scattering angles, and decreases to zero for
exactly backward scattering.  But even at $90^{\circ }$ ($y=1/2$), it is
80\%.

Note that the spin averaged square of the matrix element is
$$
   \overline{|M|^{2}} = \frac {4e^{4}q_{a}^{2}}{y^{2}}
   \left(1-y+\frac {1}{2}y^{2} \right),
\eqno(22)
$$
while the spin-dependent part is
$$
   C \overline{|M|^{2}} = \frac {4e^{4}q_{a}^{2}}{y^{2}}
   \left(1-y \right).
\eqno(23)
$$
In the structure functions, the coefficient of $1-y$ feeds into
$F_{2}$, while the coefficient of $y^{2}/2$ feeds into $F_{1}$.  Thus the
transverse spin asymmetry we are discussing can be regarded as a
100\% asymmetry in $F_{2}$ and a zero asymmetry in $F_{1}$.

\section {Predictions for DIS}
\label{dispred}

The measurements are of the single hadron distribution in the
`current quark fragmentation region' of collisions of {\em
un}polarized leptons on transversely polarized nucleons.  An
asymmetry of the cross section is expected under reversal of the
nucleon polarization as a function of the azimuthal angle $\phi $ of
the outgoing hadron around the jet axis, which is defined by parton
model kinematics.  The asymmetry will have a $\sin\phi $ dependence.

To a first approximation, the asymmetry is predicted to be the
product of the polarization of the quarks in the
initial nucleon, the spin transfer coefficient of the hard
scattering, and the analyzing power of the fragmentation.  In
principle the asymmetry is a function of all the kinematic
variables in the problem.  The effect should be largest in the
valence region for both the parton density and the fragmentation,
say for $\xbj \gsim 0.3$ and $z \gsim 0.3$.  One could even
choose the hadron to be the leading hadron in the jet.

The data should be analyzed separately for different charges, and
ideally for different flavors of hadron (e.g., kaon versus pion
versus proton).  For scattering on a proton target, the biggest
asymmetry will probably be for $\pi ^{+}$,
since the $\pi ^{+}$ has a valence
$u$ quark in common with the proton and deep inelastic scattering
dominantly goes with $u$ quarks in the valence region.  It is
possible that the asymmetry will reverse sign for $\pi ^{-}$, and the
value for $\pi ^{0}$ is not obvious.  It would also be interesting to
see how big the asymmetry is for $K^{+}$, which has a valence
$u$ quark; this would probe the difference in the creation of
strange quark-antiquark pairs in fragmentation as compared to
creation of $u$ and $d$ pairs.

Furthermore there will be a strong dependence of the asymmetry on
the transverse momentum $\kt$ of the hadron relative to the jet
axis.  This dependence will be roughly of the form of \er(15), up
to an overall coefficient, and it will be important to verify
this dependence.  At small $\kt$, there is a kinematic zero,
while, at large $\kt$, QCD predicts the asymmetry to be of higher
twist.  It is at $\kt$s of a few hundred MeV to about a GeV that
one can expect the biggest asymmetry.

Note that the unpolarized cross section itself has a significant
azimuthal dependence \ref[EMCazi].  This is predicted \ref[azi]
to be leading twist at large $\kt$, but higher twist at low
$\kt$, unlike the behavior of the transverse spin asymmetry.  But
clearly this would be a confounding effect to be careful about in
the experiments.  One must measure the actual spin asymmetry in
the cross section and not just the azimuthal dependence of
the cross section.

Precise quantitative predictions cannot be made at present since
the asymmetry depends on some not-yet-measured nonperturbative
functions.  Interesting comparisons can be made with data from
proton-proton scattering and from final state correlations in
$e^{+}e^{-}$ annihilation.  Ultimately we must wait for experiments to
give us the values of the nonperturbative functions.  They will
probe the spin and chiral structure of both the initial hadron
and of the fragmentation.  The spin and chiral structure of
fragmentation is at present a very little explored subject.

But we can suggest the general size, based on experience in other
situations.  The quarks in the hadron might be 50\%
polarized, as
for the helicity asymmetry.  The spin transfer in the hard
scattering is mostly 80\% to 100\%, from calculation.
Finally, the fragmentation could have an analyzing power
of tens of percent.  (The last figure is the most uncertain!).
Overall we can then reasonably expect about a 10\% asymmetry in
the cross section.

\section {Hadron-Hadron Scattering}
\label{hadhad}

Exactly analogous measurements can be made in jet production in
hadron-hadron scattering.  To define the jet axis, one could
measure the jet axis in a conventional manner; this would probably
have too large an uncertainty.  An example of a better way would
be to measure the correlation between the leading particles in
opposite jets. Intrinsic transverse momentum effects generate
out-of-plane transverse momentum for these particles.  The
maximum asymmetry will be when the spin vector of the incoming
polarized proton is in the plane of the hard scattering.

More theoretical analysis is needed here.

\section {Model Calculations of Fragmentation}
\label{model}

In this section I will present a very simple
model calculation of the
transversity dependent fragmentation function.  It will show
that the spin-dependence is permitted by the symmetries of QCD,
but that it requires non-trivial phases in the interaction.
Furthermore, it will be clear that the transversity dependence is
caused by the spontaneous breaking of chiral symmetry.

The model is a sigma model for pions coupled to quarks in the
manner suggested by Georgi and Manohar \ref[GM].  The Lagrangian
is to be considered an effective Lagrangian for suitable areas of
nonperturbative QCD.  While the model is not perfect, it does
contain two important features of QCD: the quark degrees of
freedom and the chiral symmetry breaking.  We will calculate the
lowest order graph for the fragmentation of a quark to a pion.
To get a transversity dependence we will see that we must dress
the propagators so that they acquire imaginary parts.

   \defpsfig{4}{2.5}{Lowest order graph for fragmentation of quark to
   pion.  The solid line is a quark and the dashed line is a
   pion.  }{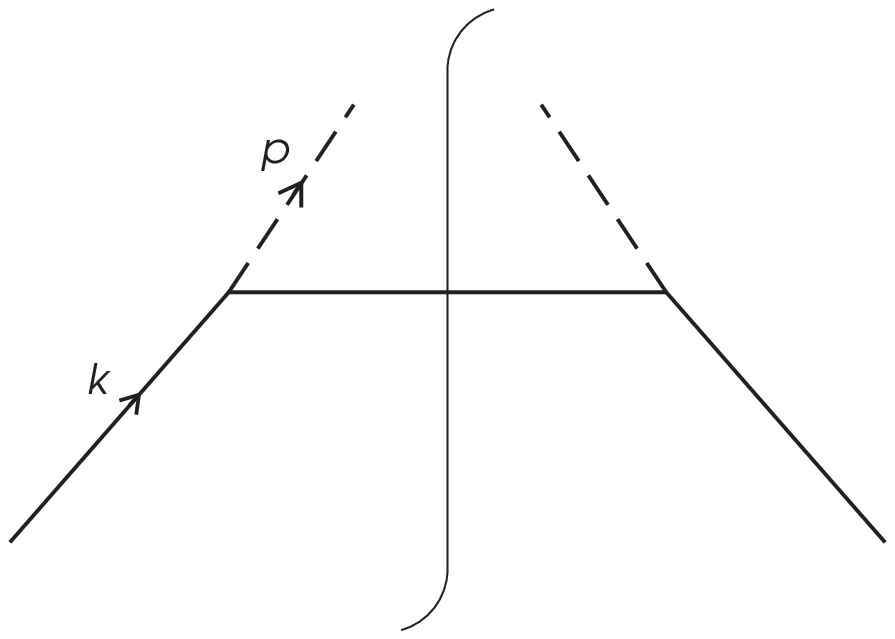}{800}{1.8}

\begin{sloppypar}
The graph is shown is \fig{4}, and its contribution to the
unpolarized fragmentation follows from the definition \eq(3):
$$\eqalign{
   \hat D(z,k_{\perp }) =
   \frac {g^{2}}{16\pi ^{4}} \int dk^{-}  \frac {1}{(k^{2}-M^{2})^{2}} &
   2\pi  \delta ^{({+})}\left( (k-p)^{2} - M^{2}\right)
\cr
  &
   \frac {1}{4} \tr \gamma ^{+} (k\llap /+M)
      \gamma _{5} (k\llap /-p\llap /+M) \gamma _{5} (k\llap /+M).
}\eqno(24)
$$
Here $g$ is the pion quark coupling, presumably around a third of
$g_{\pi NN}$, for which $g_{\pi NN}^{2}/(16\pi ^{2})\approx 1.1$.
The quark mass $M$ is due
to chiral symmetry breaking, so it should be around 300 MeV.  For
simplicity, we will neglect the pion mass.  Then a simple
calculation gives
$$
   \hat D =
   \frac {g^{2}}{16\pi ^{3}} \frac {1}{k_{\perp }^{2}+M^{2}} ,
\eqno(25)
$$
and results in a transverse momentum distribution with the
canonical 300 MeV width.  The $1/k_{\perp }^{2}$ tail at
large $k_{\perp }$ is not to be trusted:
it comes from a region where the quark is far off its mass shell,
and there the sigma model is entirely inappropriate.  Note the
flavor structure: for different charge states in the
fragmentation of a $u$-quark, we have
$u\to \pi ^{{+}}:u\to \pi ^{0}:u\to \pi ^{{-}} = 2:1:0$.

\end{sloppypar}

Similarly, the transversity dependent fragmentation function
(\eq(4)) in our model is
$$\eqalign{
  \Delta  \hat D(z,k_{\perp }) =
  \frac {g^{2}}{16\pi ^{4}} \int dk^{-} &  \frac {1}{(k^{2}-M^{2})^{2}}
  2\pi  \delta ^{({+})}\left( (k-p)^{2} - M^{2}\right)
\cr
 &
   \frac {1}{4} \tr \gamma ^{+}\gamma _{5}s_{\perp }\cdot \gamma _{\perp }
   (Ak\llap /+BM) \gamma _{5} (k\llap /-p\llap /+M)
   \gamma _{5} (A^{*}k\llap /+B^{*}M).
}\eqno(26)
$$
Here, we have anticipated the need for imaginary parts in the
amplitudes, and have used a dressed quark propagator
$i(Ak\llap /+BM)/(k^{2}-M^{2})$,
where the scalar coefficients $A$ and $B$
depend on $k^{2}$.  On-shell renormalization gives them the value 1
when $k^{2}=M^{2}$.  The result is

$$
   \Delta \hat D = \frac {g^{2}}{16\pi ^{3}}
      \frac {2 M \Im A^{*}B}{(k_{\perp }^{2}+M^{2})^{2}}
      \frac {(1-z)}{z} (s_{\perp }^{y}k^{x} - s_{\perp }^{x}k^{y}) .
\eqno(27)
$$
Hence the analyzing power is the ratio:
$$
   \frac {\Delta \hat D}{\hat D} =
      \frac {2 M \Im A^{*}B}{k_{\perp }^{2}+M^{2}} \frac {(1-z)}{z}
      (s_{\perp }^{y}k^{x} - s_{\perp }^{x}k^{y}) .
\eqno(28)
$$
This exhibits the properties claimed earlier: It gives a $\sin\phi $
dependence on the azimuth of the transverse momentum, with a
kinematic zero at $k_{\perp }=0$.
A significant analyzing power evidently
depends on the phases of $A$ and $B$ being substantially
different, but there is presumably no reason why this should not
be the case.

There is an apparent blow up of \er(28) when $z\to 0$.  This is not
actually the case, since the virtual quark then approaches its
mass shell, where $A=B=1$.

\section {Conclusions}
\label{concl}

I have proposed that one can probe the transverse polarization of
a quark that initiates a jet by measuring the azimuthal
dependence of the hadrons in the jet, especially the leading
hadrons.  There are certainly other processes than the two
explained in this paper that could be exploited.  The typically
large spin transfer in the hard scattering will aid in generating
large asymmetries.

Perhaps the most interesting aspect of the idea is that it gives
one a direct handle on how the chiral properties of the strong
interactions couple to fragmentation.  In particular, it would
appear from the model calculations in \secref{model} that the
fragmentation asymmetry will only appear if there is breaking of
chiral symmetry.  Thus, in the context of fragmentation, it
probes the known spontaneous breaking of chiral symmetry.

Conceivably, the measured asymmetry could be very small.
This would require that the quarks in a transversely
polarized proton have a small polarization or that the
analyzing power of the fragmentation is small.  The first
possibility seems highly unlikely, given what we know of the
other parton distributions in the valence region. The model
calculation for the fragmentation indicates that there is no
symmetry reason why the analyzing power should be small.
Perhaps the best suggestion that a significant analyzing
power is likely comes from the frequency of large spin
asymmetries that are experimentally measured in
hadron-induced processes.

The differences between the longitudinal and the transverse
spin asymmetries in the distribution function are sensitive to
relativistic and spin-orbit effects in the proton wave function,
and are thus of intrinsic interest.


\section *{Acknowledgments}

I would like to thank several colleagues for useful
conversations, in particular Bob Carlitz, Steve Heppelmann,
Bob Jaffe, Glenn Ladinsky, Lech Mankiewicz, Jian-Wei Qiu and
George Sterman. This work was supported in part by the U.S.
Department of Energy under grant DE-FG02-90ER-40577 and by
the Texas National Laboratory Research Commission.


\section*{Appendix}

Sivers \ref[Sivers] suggested that the $\kt$ distribution of a quark
in a hadron could have an azimuthal asymmetry when the
initial hadron has transverse polarization.  However, as we
will now show, such an asymmetry is prohibited because QCD
is time-reversal invariant.

Consider
\begin{eqnarray*}
   \hat f_{a/A}(x,\kt;\alpha ,\alpha ') &\equiv &
\\ &&\hspace*{-1cm}
   \int  \frac {\d y^{-} \d^{2}y_{\perp }}{(2\pi )^{3}}
   \e^{-ixp^{+}y^{-}+i\kt\cdot y_{\perp }}
   \langle p\alpha | \, \bar \psi _{a}(0,y^{-},y_{\perp }) \,
   \frac {\gamma ^{+}}{2} \, \psi _{a}(0) \, |p\alpha '\rangle ,
\end{eqnarray*}
which is a matrix in the helicity indices $\alpha $ and $\alpha '$. This
quantity is the same as eq.\ (1), except that we have
replaced the hadron states by particular spin states in a
helicity basis. ($\alpha $ and $\alpha '$ take the values $L$ or $R$.)
Contracting this matrix with the spin density matrix for the
incoming hadron gives eq.\ (1).

It can readily be checked that the matrix $\hat
f_{a/A}(x,\kt;\alpha ,\alpha ')$  is hermitian, i.e.,
$$
    \left[\hat f_{a/A}(x,\kt;\alpha ,\alpha ')\right]^{\dagger }
    \equiv  \left[\hat f_{a/A}(x,\kt;\alpha ',\alpha )\right]^{*}
    = \hat f_{a/A}(x,\kt;\alpha ,\alpha ').
$$

Furthermore, let us choose coordinates in which the incoming
hadron has zero transverse momentum, $p_{\perp }=0$.  Then invariance
under rotations about the $z$ axis shows that the diagonal
elements, $\hat f_{a/A}(x,\kt;RR)$ and $\hat f_{a/A}(x,\kt;LL)$
are functions of the length but not the angle of $\kt$.
Parity invariance implies that these two diagonal elements
are equal:
$$
   \hat f_{a/A}(x,\kt;RR)=\hat f_{a/A}(x,\kt;LL).
$$

We now apply a time-reversal transformation followed by a
parity inversion.  With the Bjorken and Drell \ref[BD]
conventions for the Dirac matrices, we have
$$
    (\PT)^{\dagger } \psi _{\alpha }(x) \PT = PT \psi (-x),
$$
with $PT=i\gamma ^{0}\gamma ^{1}\gamma ^{3}$.  Note that momenta are unchanged
but
helicity states get reversed under $\PT$.  Moreover there is
a relative sign between the transformation of left and
right-handed helicities:
\begin{eqnarray*}
    \PT|p,L\rangle  &=& |p,R\rangle ,
\\
    \PT|p,R\rangle  &=& -|p,L\rangle ,
\end{eqnarray*}
where we have ignored an overall phase.  The relative minus
sign is essential, in order that $\TOp^{2}$ be $-1$ when acting
on fermionic states.  This sign can be verified from the
transformation law of the Dirac field provided that one
remembers that $\TOp$ is antilinear and that $T^{*}=-T$.

It follows that
\begin{eqnarray*}
   \hat f_{a/A}(x,\kt;LR) &=&
   -  \int  \frac {\d y^{-} \d^{2}y_{\perp }}{(2\pi )^{3}}
   \e^{-ixp^{+}y^{-}+i\kt\cdot y_{\perp }}
\\ && \hspace{1cm}
   \langle p,R| \, \TOp^{\dagger }\bar \psi _{a}(0,y^{-},y_{\perp }) \TOp\,
   \frac {\gamma ^{+}}{2} \, \TOp^{\dagger }\psi _{a}(0)\TOp \, |p,L\rangle
^{*}
\\
 &=&
   -  \int  \frac {\d y^{-} \d^{2}y_{\perp }}{(2\pi )^{3}}
   \e^{-ixp^{+}y^{-}+i\kt\cdot y_{\perp }}
\\ && \hspace{1cm}
   \langle p,R| \, \bar \psi _{a}(0,-y^{-},-y_{\perp }) \,
   PT \frac {\gamma ^{+}}{2} PT \, \psi _{a}(0) \, |p,L\rangle ^{*}
\\
   &=& - \hat f_{a/A}(x,\kt;LR),
\end{eqnarray*}
where the complex conjugation arises because the time
reversal operator $\TOp$ is antilinear, and the last line
follows from the hermiticity of $\hat f(\alpha \alpha ')$.  It is now
immediate that the off-diagonal elements are zero:
$$
    \hat f(LR) = \hat f(RL) =0.
$$

The matrix $\hat f(\alpha \alpha ')$ is therefore proportional to the
unit matrix, so that there is no dependence of the
transverse momentum distribution on the spin of the incoming
hadron. This contradicts Sivers' [Sivers] suggestion.

%
%
\begin {thebibliography}{ABCD}
\defref{AM}{X. Artru and M. Mekhfi, \zphc \rf{45}{669}{90}.  }
\defref{azi}{J. Chay, S.D. Ellis and W.J. Stirling, \pr
   \rf{D45}{46}{92}.  }
\defref{BB}{I. Balitsky and V. Braun, \np \rf {B361}{93}{91}.}
%
\defref{BD}{J.D. Bjorken and S.D. Drell, ``Relativistic
   Quantum Fields'' (McGraw-Hill, New York, 1965).  }
\defref{CarlitzDY}{R. Carlitz and R. Willey, ``Single
   Spin Asymmetries in Muon Pair Production'' in Proceedings
   of the Polarized Collider Workshop, ed.\ J.C. Collins, S.
   Heppelmann and R. Robinett, AIP Conference
   Proceedings 223 (AIP, New
   York, 1991);
   R. Carlitz and R. Willey, ``Determining
   Gluon And Anti-Quark Spin Fractions Via Single Spin
   Asymmetry Measurements'', preprint PITT-91-11;
   R. Carlitz and R. Willey, \pr \rf{D45}{2323}{92}.}
\defref{CS}{J.C. Collins and D.E. Soper, \np \rf{B194}{445}{82},
   and references therein.}
\defref{CSback}{J.C. Collins and D.E. Soper, \np
   \rf{B193}{381}{81}.}
%
\defref{CSS}{J.C. Collins, D.E. Soper and G. Sterman,
   ``Factorization of Hard Processes in QCD''
   in ``Perturbative QCD" (A.H. Mueller, ed.) (World
   Scientific, Singapore, 1989), and references therein.  }
\defref{CSS1}{J.C. Collins, D.E. Soper and G. Sterman, \np
   \rf{B261}{104}{75} and \rf{B308}{833}{88};
   G.T. Bodwin, \pr \rf {D31}{2616}{85} and \rf{D34}{3932}{86}.}
%
\defref{CSS2}{J.C. Collins, D.E. Soper and G. Sterman, \np
   \rf{B250}{199}{85}.  }
%
\defref{DG}{J.T. Donohue and S. Gottlieb, \pr
   \rf{D23}{2577,2581}{81}.  }
%
\defref{DMG}{R.H. Dalitz, G.R. Goldstein, and R. Marshall,
   \zphc \rf{42}{441}{89}, and \pl \rf{B215}{783}{88};
   G.R.
   Goldstein, ``Determining Quark Helicity from Jet
   Distributions'' in Proc.\ of 8th Int.\ Symp.\ on High
   Energy Spin Physics, Minneapolis, MN,
   ed.\ K.J. Heller, AIP Conference Proceedings 187
   (AIP, New York, 1989) }
\defref{DYJCC}{J.C. Collins, \pr\rf{D21}{2962}{80}.  }
%
\defref{Einhorn}{M.B. Einhorn, in
   ``Spin Correlations in Quark and Gluon Fragmentation'' in
   Proc.\ of 8th Int.\ Symp.\ on High Energy Spin Physics,
   Minneapolis, MN, ed.\ K.J. Heller,
   AIP Conference Proceedings 187 (AIP, New York,
   1989). }
\defref{EMCazi}{European Muon Collaboration, M Arneodo et al.,
   \zphc \rf{34}{277}{87}.}
%
\defref{EMCfrag}{European Muon Collaboration, M Arneodo et al.,
   \np \rf{B321}{541}{89}.}
%
\defref{EMT}{A.V. Efremov, L. Mankiewicz and N.A. T\"ornqvist,
  \pl\rf{B284}{394}{92}.}
%
\defref{F}{R.P. Feynman, ``Photon-Hadron Interactions''
   (Benjamin, Reading MA, 1972).  }
\defref{Gas}{S. Gasiorowicz, ``Elementary Particle Physics''
   (Wiley, New York, 1966), page 515.  }
\defref{GM}{A. Manohar and H. Georgi, \np \rf{B234}{189}{84}.}
%
\defref{GW}{R. Gastmans and T.T. Wu, ``The Ubiquitous Photon''
   (Oxford University Press, Oxford, 1990).  }
%
%
\defref{JJ}{R.L. Jaffe and X.-D. Ji, \prl \rf{67}{552}{91};
   X.-D. Ji, \pl \rf{B284}{137}{92}.  }
%
\defref{LC}{D.E. Soper, \pr \rf{D15}{1141}{77};
   C. Bouchiat, P. Fayet and P. Meyer, \np \rf{B34}{157}{71}.  }
\defref{MOS}{R. Meng, F. Olness, and D.E. Soper, \np
   \rf{B371}{79}{92}.  }
\defref{Nacht}{O. Nachtmann, \np \rf{B127}{314}{77}.}
\defref{pion}{E704 Collaboration: D.L. Adams et al.,
   \pl \rf{B264}{462}{91},
   \pl \rf{B261}{201}{91}, \pl \rf{B276}{531}{92}, and
   references quoted there to earlier experiments at lower
   energies.}
%
\defref{polfact}{J.C. Collins, ``Hard Scattering in QCD with
   Polarized Beams'', Penn State preprint PSU/TH/100.  }
\defref{QS}{J.-W. Qiu and G. Sterman, \np\rf{B353}{105, 137}{91} and
   \prl\rf{67}{2264}{91}. }
%
\defref{RS}{J.P. Ralston and D.E. Soper, \np \rf{B152}{109}{79}.  }
\defref{Sivers}{D. Sivers, \pr \rf{D41}{83}{90} and \rf{D43}{261}{91}.
}
\defref{Tk}{F.V.\ Tkachov,
   ``Euclidean Asymptotic Expansions of Green Functions of
   Quantum Fields. (I) Expansions of Products of Singular
   Functions'', preprint FERMILAB-PUB-91/347-T,
   Int.\ J. Mod.\ Phys.\ A (1992) (in print) and refs.\ therein;
   G.B.\ Pivovarov and F.V.\ Tkachov: ``Euclidean Asymptotic
   Expansions of Green Functions of Quantum Fields. (II)
   Combinatorics of the $As$-operation", preprint
   FERMILAB-PUB-91/345-T;
   Int.\ J. Mod.\ Phys.\ A (1992) (in print).}
\defref{trfrag}{R. Carlitz, J.C. Collins, S. Heppelmann, R.
   Jaffe, X. Ji, G. Ladinsky,
   ``Measuring Transversity Densities in Singly Polarized
   Hadron-Hadron Collisions'', Penn State preprint
   PSU/TH/101, in preparation.  }
\defref{trjet}{K. Hidaka, E. Monsay and D. Sivers,
   \pr\rf{D19}{1503}{79}.}
%
\end {thebibliography}

\end {document}